\documentclass[12pt,twoside,fleqn]{article}
\usepackage{epsfig}
\usepackage{multirow}
\usepackage{exscale}
\def\lsim{\raise0.3ex\hbox{$<$\kern-0.75em\raise-1.1ex\hbox{$\sim$}}}
\def\gsim{\raise0.3ex\hbox{$>$\kern-0.75em\raise-1.1ex\hbox{$\sim$}}}

\newcommand{\beqn}{\begin{equation}}
\newcommand{\eqn}{\end{equation}}
\newcommand{\bqa}{\begin{eqnarray}}
\newcommand{\eqa}{\end{eqnarray}}
\newcommand{\bqas}{\begin{eqnarray*}}
\newcommand{\eqas}{\end{eqnarray*}}
\newcommand{\bdm}{\begin{displaymath}}
\newcommand{\edm}{\end{displaymath}}

\newcommand{\ie}{{\sl i.e.~\/}} 

\begin{document}
\thispagestyle{empty}
\mbox{} \hfill BI-TP 2001/28\\
\mbox{} \hfill DESY 01-175\\
\mbox{} \hfill revised 
\begin{center}
{{\large \bf A Lattice Calculation of Thermal Dilepton Rates}
 } \\
\vspace*{1.0cm}
{\large F. Karsch$^1$, E. Laermann$^1$, P. Petreczky$^1$, S. Stickan$^1$ 
\\[2mm]
and 
\\[2mm]
I. Wetzorke$^2$}

\vspace*{1.0cm}
{\normalsize
$\mbox{}$ {$^1$ Fakult\"at f\"ur Physik, Universit\"at Bielefeld,
D-33615 Bielefeld, Germany}

$\mbox{}$ {$^2$ NIC/DESY Zeuthen, Platanenallee 6, D-15738 Zeuthen,
Germany}
}
\end{center}
\vspace*{1.0cm}
\centerline{\large ABSTRACT}

\baselineskip 20pt

\noindent
Using clover improved Wilson fermions we
calculate thermal vector meson correlation 
functions above the deconfinement phase transition 
of quenched QCD. At temperatures $1.5\; T_c$ and 
$3\; T_c$ they are found to differ by less than
15\% from that of a freely propagating quark
anti-quark pair. This puts severe constraints
on the dilepton production rate and in particular
rules out a strong divergence of the dilepton
rate at low energies. The vector spectral 
function, which has been reconstructed using the
Maximum Entropy Method, yields an enhancement of
the dilepton rate over the Born rate of at most a factor
two in the energy interval $4\; \lsim\; \omega/T\; \lsim\; 8$ 
and suggests that the spectrum is cut-off at low energies
by a thermal mass threshold of about $(2-3)\; T$.

\vfill
\noindent
\eject
\baselineskip 15pt

\noindent

\section{Introduction}

The dilepton spectrum is one of the key observables in the study
of thermal properties of the hot and dense medium created in
heavy ion collisions \cite{Kapusta,Toimela,basics}. 
Thermal modifications of the 
dilepton spectrum have been observed at large invariant masses
where they reflect the suppression of heavy quark resonances
\cite{jpsi}. At low energies it is expected that non-perturbative
in-medium modifications of the quark anti-quark interactions 
as well as quark anti-quark annihilation in the quark gluon plasma
influence the thermal dilepton rate. Perturbative 2-loop
\cite{Aurenche} {calculations as well as calculations based on hard 
thermal loop (HTL) resummed perturbation theory } \cite{htl} 
lead to an enhancement of the spectrum at low energies,
which dominates over the suppression arising from massive quasi-particle
poles \cite{Kapusta,htl} or non-perturbative gluon condensates \cite{Mustafa}.
Indeed, an enhancement of low mass dilepton production has been observed 
in heavy ion experiments \cite{ceres}. 

The thermal dilepton rate describing the production of lepton pairs
with energy $\omega$ and total momentum $\vec{p}$ is related to the
Euclidean correlation function of the vector current, 
$J_V^{\mu}(\tau, \vec{x}) = \bar{\psi}(\tau, \vec{x}) \gamma_\mu 
\psi(\tau, \vec{x})$, which can be calculated numerically in the 
framework of lattice regularized QCD. 
This relation is established through the vector spectral function, 
$\sigma_V(\omega,\vec{p},T)$, which directly gives the differential 
dilepton rate in two-flavor QCD,
\begin{equation}
{{\rm d} W \over {\rm d}\omega {\rm d}^3p} = 
{5 \alpha^2 \over 27 \pi^2} {1\over \omega^2 ({\rm e}^{\omega/T} - 1)}
\sigma_V(\omega,\vec{p},T) \quad ,  
\label{rate}
\end{equation}
and at the same time is related to the Euclidean vector meson correlation 
function, $G_V (\tau, \vec{p},T) = \int {\rm d}^3 x \exp (i\vec{p}\cdot\vec{x})
\langle J_V^{\mu}(\tau, \vec{x}) J_{V\mu}^\dagger(0,\vec{0})\rangle$, 
through an integral equation,
\begin{equation} 
G_V (\tau, \vec{p},T) = \int_0^{\infty} \hspace*{-0.2cm}{\rm d} \omega \;
\sigma_V(\omega,\vec{p},T)\;
{ {\rm ch} (\omega(\tau -1/2T)) \over {\rm sh} (\omega/2T)}~, 
\label{correlator}
\end{equation}
where the Euclidean time $\tau$ is restricted to the interval $[0,1/T]$.
A direct calculation of the differential dilepton rate thus becomes 
possible, if the above integral equation can be inverted
to determine $\sigma_V(\omega,\vec{p},T)$. Although finite 
temperature lattice calculations, which are performed on lattices with
finite temporal extent $N_\tau$, will generally provide information on 
$G_V (\tau, \vec{p},T)$ only for a discrete and finite set of Euclidean 
times $\tau T = k/N_\tau~,~k=1,...,N_\tau$, this may be achieved 
through an application of the maximum entropy method (MEM) 
\cite{Hatsuda,Wetzorke}. We will present first results on such a 
calculation in this letter. However, even without invoking statistical
tools like MEM accurate numerical results on $G_V(\tau,\vec{p},T)$ will
themselves provide stringent constraints on spectral functions 
calculated in other perturbative or non-perturbative approaches. This 
in turn will put constraints on the thermal dilepton rates.

We present here results of a calculation of vector current
correlation functions 
and reconstructed spectral functions at
two temperatures above $T_c$, \ie $T= 1.5\; T_c$ and $3\; T_c$. 
We will restrict our discussion to the 
zero momentum limit ($\vec{p}=0$) of these quantities and we will
therefore suppress any explicit momentum dependence in the arguments
of $G_V$ and $\sigma_V$ in the following.

\section{Thermal correlation functions}

The correlation functions $G_V(\tau,T)$ have been calculated within the 
quenched
approximation of QCD using non-perturbatively improved clover  
fermions \cite{clover,luescher}. 
This eliminates the ${\cal O}(a)$
discretization errors in the fermion sector. 
Moreover, in order to explicitly
control the cut-off dependence of the numerical results we have used
lattices of size $N_\sigma^3\times N_\tau$ with fixed ratio of spatial 
to temporal lattice extent,
$N_\sigma / N_\tau = 4$, and two different temporal lattice sizes, 
$N_\tau =12$ and 16, respectively. Due to the large temporal extent
of the lattice, the lattice spacing is rather small already at the 
deconfinement transition and becomes even smaller for larger 
temperatures.
Using $T_c = 265$~MeV \cite{Boyd} to set a scale for the lattice 
spacing one finds,
$a=1/N_\tau T \simeq (0.75 T_c/T\; N_\tau)$~fm. 
For the two different temperatures
and lattice sizes used here it ranges from $a= 0.015$~fm to $a= 0.04$~fm. 
The comparison of results obtained on lattices with different temporal 
extent and fixed $T$ indeed confirms that the results for $G_V(\tau, T)$ 
are not significantly influenced by finite cut-off effects. 

\begin{table}[t]
\begin{center}
\begin{tabular}{|c|c|c|c|c|c|c|c|}
\hline
{$T/T_c$}&$N_\tau$&$\beta$ &$c_{sw}$&$\kappa_c$& $m_q a$ &
            $Z_V$&\#~conf. \\ \hline
1.5& 12& 6.640 & 1.4579 & 0.13536 &           & 0.8184 &25 \\ 
  ~& 16& 6.872 & 1.4125 & 0.13495 & 0.0014(2) & 0.8292 &40 \\ \hline
3.0& 12& 7.192 & 1.3550 & 0.13440 & 0.0031(4) & 0.8421 &40 \\ 
  ~& 16& 7.457 & 1.3389 & 0.13390 & 0.0033(2) & 0.8512 &40 \\ 
\hline
\end{tabular}
\end{center}
\caption{Simulation parameters for the calculation of $G_V(\tau, T)$ 
on lattices of size $48^3\times 12$
and $64^3\times 16$. The third column gives the gauge coupling
$\beta =6/g^2$. 
The values used for the clover coefficient $c_{sw}$, the critical
hopping parameter $\kappa_c$ and the current renormalization
constant $Z_V$ are discussed in the text. The values for $T/T_c$
given in the first column are based on results for the non-perturbative
$\beta$-function given in Ref.\cite{Boyd}. 
}
\label{tab:statistics}
\end{table}

The thermal correlation functions have been constructed from 
quark propagators obtained from the inversion of the 
clover improved Wilson fermion matrix.
As we have performed calculations in the high temperature, chirally
symmetric phase of QCD there are no massless Goldstone modes which 
could interfere with the convergence of the iterative solvers for this matrix 
inversion. We thus could perform calculations directly in the limit of 
vanishing quark masses, \ie directly at the critical values of
the hopping parameter $\kappa_c$ given in Tab.~\ref{tab:statistics}.
The values for $c_{sw}$ as well as critical
$\kappa$ values were obtained from \cite{luescher} and interpolation.
Note that a finite temperature critical $\kappa$ defined
by a vanishing quark mass might differ from the $T=0$ value
$\kappa_c$ by finite lattice spacing and finite volume
effects. 
Indeed, the values for the quark mass obtained from the
PCAC relation \cite{Bocchicchio}
are very small but not exactly zero, 
see Tab.~\ref{tab:statistics}.
However, since 
the correlation functions above $T_c$ are almost
quark mass independent close to the chiral limit \cite{pschmidt} it is 
not crucial to hit $\kappa_c(T \neq 0)$ precisely.

Some care has to be taken over the proper renormalization 
of the vector current $J_{V \mu}$
used in the lattice calculations.
In order to be able to compare results obtained with different 
$N_\tau$ and $T$ as well as with calculations performed in different 
continuum regularization schemes we use 
the renormalized local current 
$J_{V \mu}^{\rm renorm} \equiv (2\; \kappa\; Z_V) J_{V \mu}$. 
Here 
$Z_V(g^2)$ is the current renormalization constant
for which we use the
non-perturbative values determined in
\cite{luescher}.
The overall error on the 
normalization \cite{luescher,Guagnelli}, including the
effect of the quark mass not being exactly zero
has been estimated to be less than 1 \%.

\begin{figure*}
\epsfig{file=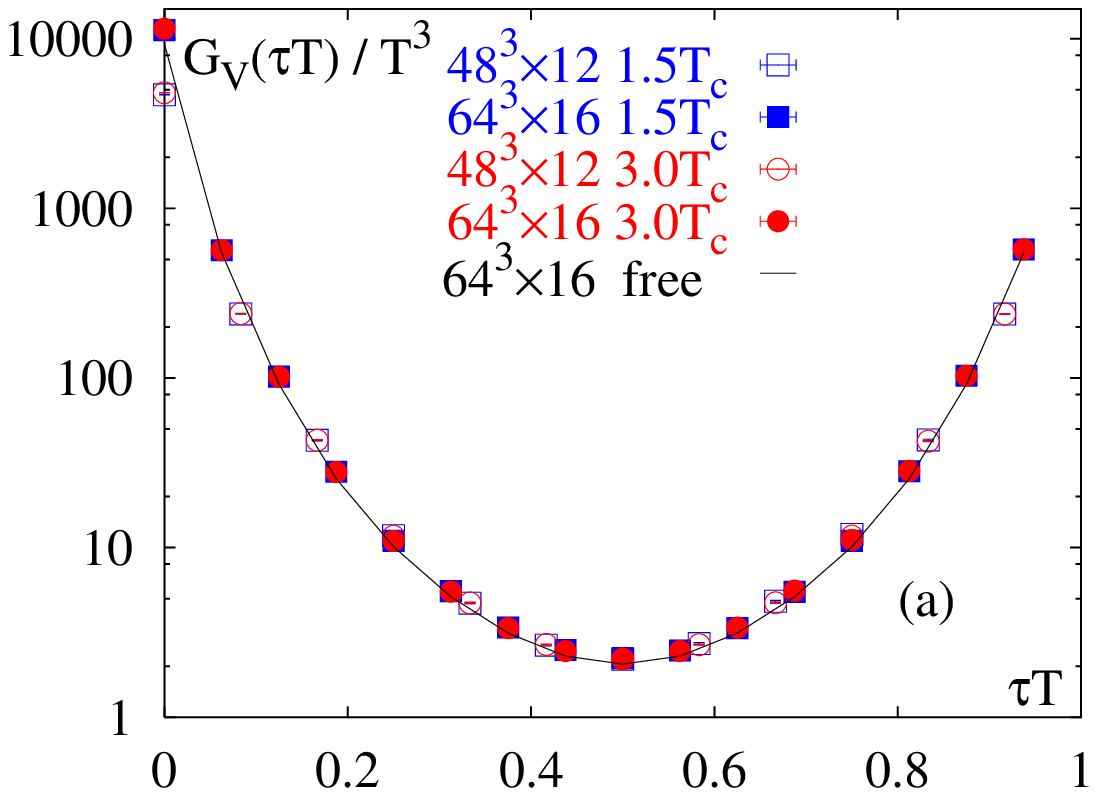,width=75mm}
\hspace{0.3cm}\epsfig{file=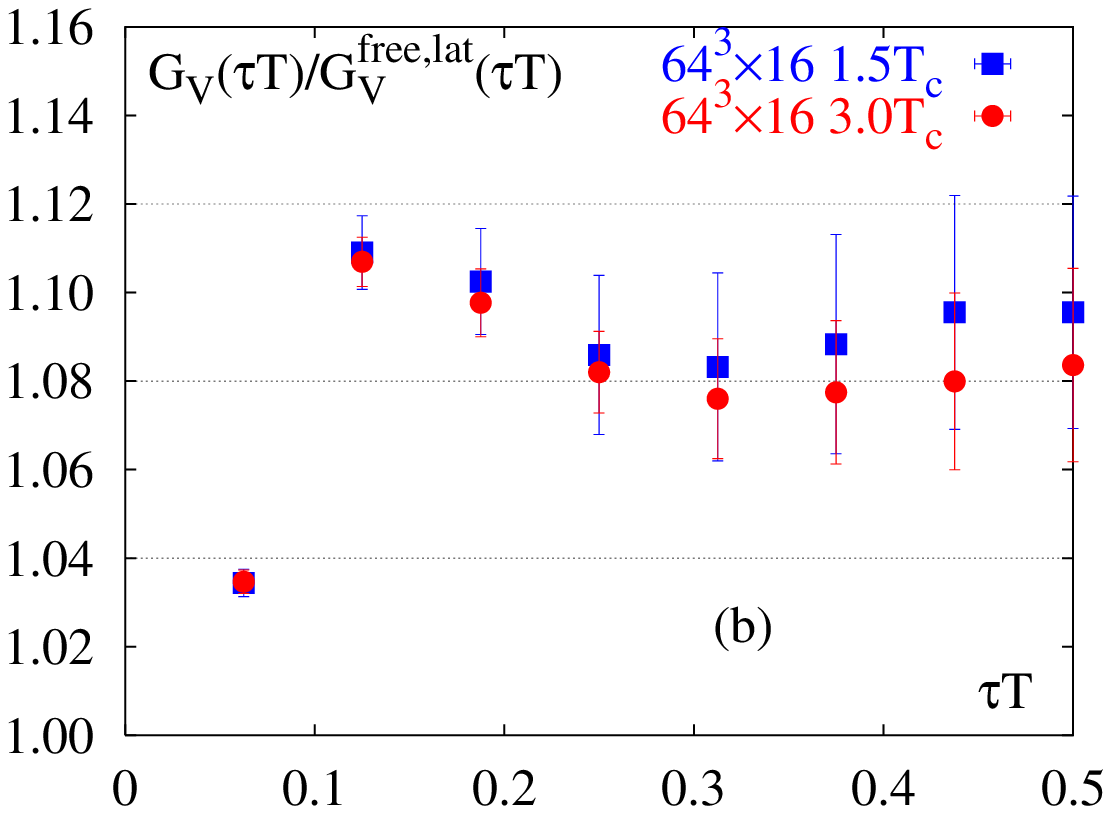,width=75mm}
\caption{
\label{fig:vector}
Thermal vector current correlation functions, $G_V$, versus Euclidean
time $\tau$ (a) and the ratio of $G_V$ and the free correlator
$G_V^{free,lat}$ (b). 
Shown are results for $T/T_c = 1.5$  and 3. The solid line in (a) shows
the  correlation function for a freely propagating quark anti-quark pair,
$G_V^{free,lat}$, calculated with Wilson fermions on a finite lattice 
with spatial extent $N_\sigma = 64$ and temporal extent $N_\tau =16$.}
\end{figure*}

In the high temperature limit the ratio $G_V/T^3$ 
can be calculated perturbatively. The leading order result is obtained 
by using free,
massless quark propagators in the calculation of the current-current
correlation function.  For vanishing momentum of the quark anti-quark 
pair ($\vec{p} = 0$) the corresponding spectral
function is given by $\sigma_V(\omega, T) = 0.75\; \pi^{-2}
\omega^2 \tanh (\omega /4T)$. In this limit the differential 
dilepton rate reduces to the Born rate,
\begin{equation}
{{\rm d} W^{\rm Born} \over {\rm d}\omega {\rm d}^3p}(\vec{p}=0) = 
{5 \alpha^2 \over 36 \pi^4} {1\over ({\rm e}^{\omega/2T} + 1)^2}
\quad .  
\label{born}
\end{equation}
In Fig.~\ref{fig:vector}a
we show our numerical results for $G_V / T^3$ calculated at 
$T/T_c = 1.5$ and 3. 
These results are compared to the lattice version of the free vector 
correlation function, $G_V^{\rm free,lat}/T^3$ \cite{free}. 
It is evident that the correlators obtained at two different temperatures 
agree with each other quite well and stay close to the leading order 
perturbative (free) vector correlation function already quite close to $T_c$. 
The deviations from $G_V^{\rm free,lat}/T^3$
are better visible in the ratio
$G_V / G_V^{\rm free,lat}$ which is shown in  Fig.~\ref{fig:vector}b.
It is noteworthy that $G_V(\tau)/T^3$ is definitely larger than the 
corresponding free correlator, $G_V^{\rm free,lat}/T^3$ for both temperatures 
and all Euclidean times $0 < \tau T < 1$. 
This is furthermore supported by a calculation of
these ratios for different values of the cut-off, e.g. for $N_\tau = 12$
and 16 at the same value of $T/T_c$. Results agree within
statistical errors. 

As is obvious from Eq.~\ref{correlator} the
correlation function $G_V (\tau, T)$ receives contributions 
from all energies $\omega$. Due to the presence of the integration kernel
$K(\tau,\omega)= 
{ {\rm ch} (\omega(\tau -1/2T)) / {\rm sh} (\omega/2T)}$
large energies are, however, exponentially suppressed.
This suppression becomes more efficient with increasing Euclidean
time $\tau T$. For $\tau T \simeq 1/2$ the correlation function thus is 
most sensitive to the low energy part of $\sigma_V(\omega,T)$.
In fact, using the free spectral function in Eq.~\ref{correlator}
we estimate that energies larger than $\omega/T \simeq 15$ contribute
only 2\% to the central value $G_V (1/2T,T)/T^3$
whereas in the low energy regime,
$\omega /T\; \lsim\; 3$, 
the corresponding contribution still amounts to 12\%.
$G_V (1/2T,T)/T^3$ thus is sensitive to modifications of the vector 
spectral functions up to $\omega/T \simeq 15$ and provides a stringent 
test of models and approximate calculations of the low energy part of 
the vector spectral function. We find
\begin{equation}
{G_V (1/2T, T) \over T^3} = \cases{
2.23 \pm 0.05 &, $T/T_c = 1.5$ \cr
2.21 \pm 0.05 &, $T/T_c = 3$}
\quad , 
\label{cor2T}
\end{equation}
which is about 10\% larger than the free, infinite volume  
value $G_V^{\rm free}(1/2T, T) \equiv 2$ as well as the corresponding 
free value on a finite $64^3\times 16$ lattice, $G_V^{\rm free, lat}
(1/2T, T) \equiv 2.0368 $.
Our result thus suggests that at least for a significant range of 
energies the vector spectral function is enhanced over the free case. 
Such an increase cannot easily be incorporated in
simple quasi-particle pictures, which only include 
massive quasi-particle poles in the quark spectral functions. They lead
to a reduction of $G_V(\tau,T)$ relative to $G_V^{\rm free} (\tau,T)$ and
in turn to a reduced differential dilepton rate 
\cite{Kapusta,Mustafa,Kar01}. 
On the other hand, HTL-resummed perturbative calculations, which incorporate 
not only poles at non-zero quasi-particle masses but also
additional contributions from cuts in
the HTL-resummed quark spectral functions, show that both
non-perturbative features of quarks propagating in a hot plasma play an 
important role for the low energy part of the meson spectral functions 
\cite{htl}. 
The cut contributions in the HTL approach \cite{htl} as well as in 2-loop
perturbative calculations \cite{Aurenche} even lead to 
divergent vector spectral functions at small energies which also 
makes $G_V(\tau,T )$ to diverge for all Euclidean times $\tau$.
Assuming that at our aspect ratio of $TV^{1/3} = 4$ finite volume effects 
do not play an important role, the mere fact that we do find a finite result
for the vector correlator via Eq.~\ref{correlator} implies that 
$\sigma_V(\omega, T)$ vanishes in the limit $\omega\rightarrow 0$.

\section{Reconstructed spectral functions}

The correlation functions $G_V$ shown in Fig.~\ref{fig:vector}
clearly stay close to the leading order perturbative result.
This suggests that also the spectral functions $\sigma_V$ are close
to that of the free case. We have reconstructed 
$\sigma_V$ from
$G_V$ using MEM, which has successfully been applied at $T=0$
\cite{Hatsuda,Yamazaki} and also 
has been tested at finite temperature \cite{Wetzorke}. In order to
take into account 
finite lattice cut-off effects in the reconstruction
of the spectral function 
we introduce a lattice version, $K_L(\tau,\omega,N_\tau)$, 
of the continuum kernel $K(\tau,\omega)$ which
appears in Eq.~\ref{correlator}. 
With the lattice kernel
the spectral function 
$\sigma_V$ for vanishing momentum is defined through 
\begin{equation}
\label{speccora}
G_V(\tau, T) =  \int_{0}^{\infty} {\rm d} \omega\;
\sigma_V (\omega,T)\; K_{L}(\tau,\omega, N_\tau)~~, 
\end{equation} 
where $K_{L}(\tau,\omega, N_\tau)$ is the finite lattice approximation 
of a kernel appropriate to describe the correlation function of a  
free boson at finite temperature, 
\begin{equation}
K_{L} (\tau,\omega,N_\tau) 
= {2\omega \over T} \sum_{n=0}^{N_\tau -1}
{\exp (- i 2\pi n\tau T) \over 
(2 N_\tau \sin(n\pi /N_\tau ))^2 + (\omega/T)^2} 
\quad.
\label{kernel}
\end{equation}
This lattice kernel differs from the $T=0$ version
used in \cite{Hatsuda} in so far as we have explicitly taken into account 
the finite temporal extent of the lattice. Of course, $K_L$ approaches
$K$ in the limit of large $N_\tau$. Moreover, for the energy range important
for the description of the correlation functions close to $\tau T=1/2$,
\ie for $\omega /T < 15$, they differ by less than 2\%
already on lattices with temporal extent $N_\tau=16$. 
Using this lattice kernel we have checked that the MEM analysis of
correlation functions constructed from given input spectral functions
successfully reproduces these input functions. In particular, we find
that the free vector spectral function for massless quarks 
can be reproduced already by using only 16 equally spaced points
in the time interval $[0,1/T]$. 

\begin{figure*}
\vspace*{0.4cm}
\hspace*{-0.6cm}\epsfig{file=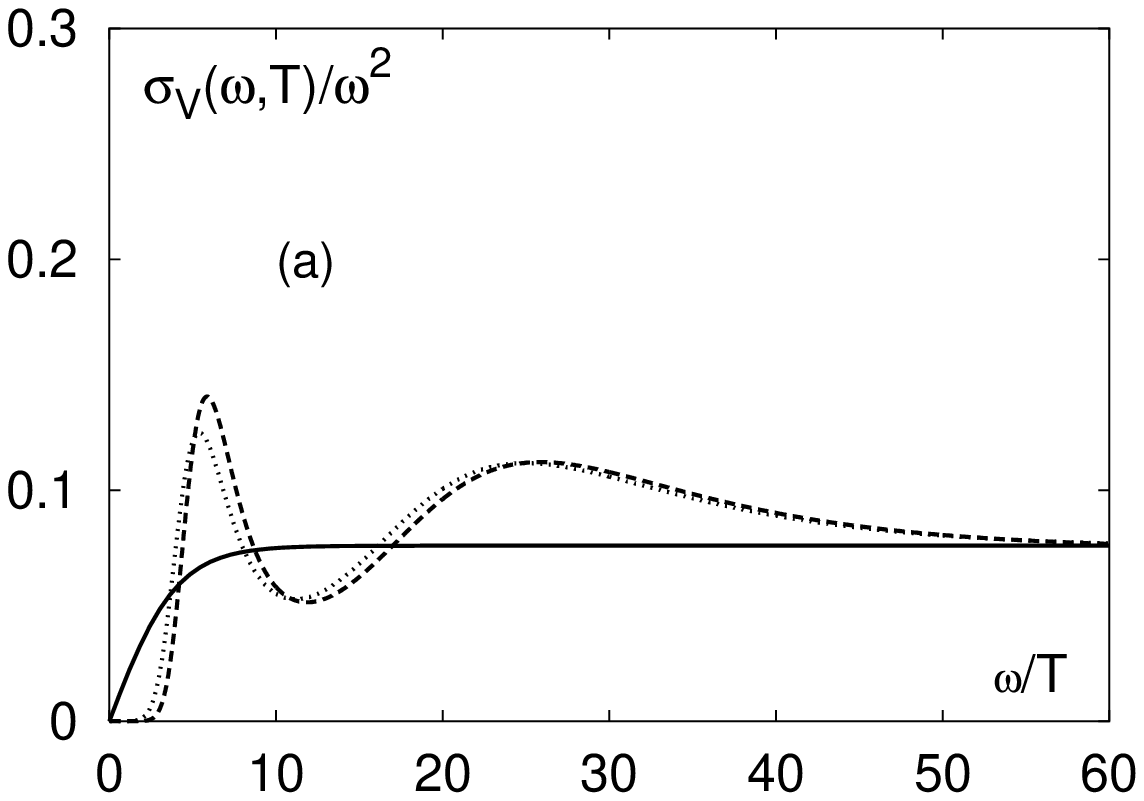,width=80mm}
\hspace*{-0.4cm}\epsfig{file=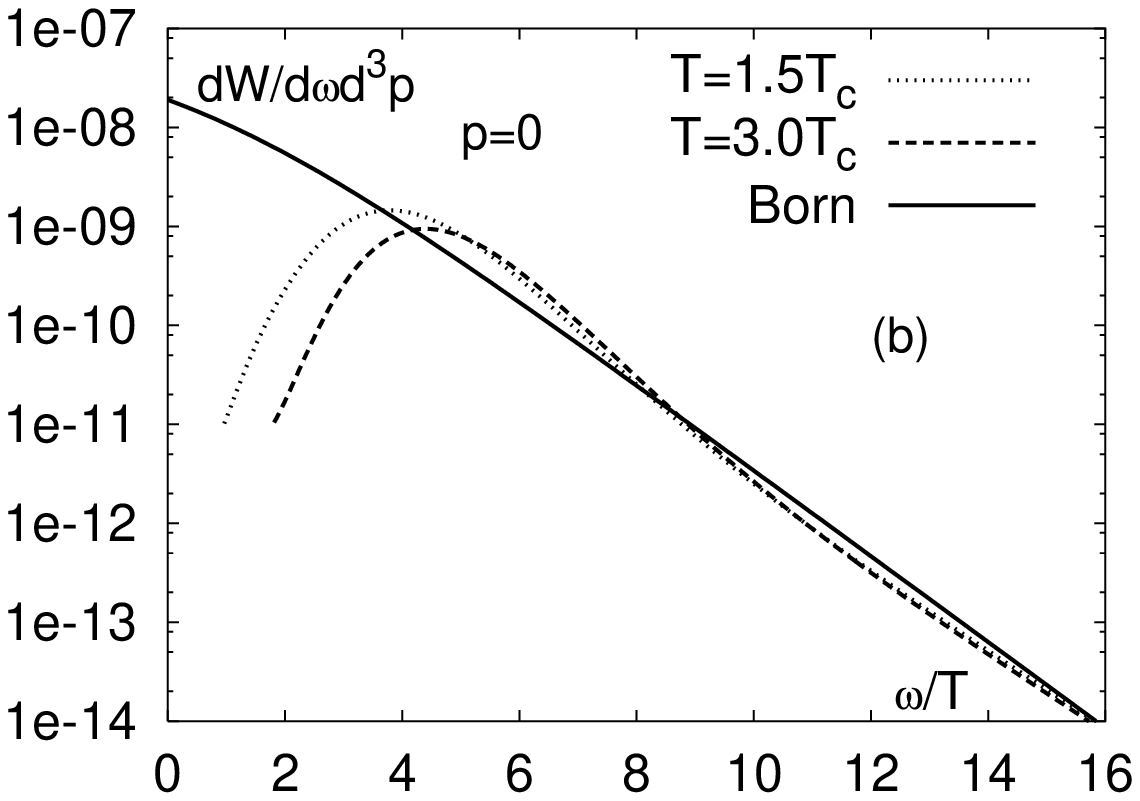,width=80mm}

\setlength{\unitlength}{1cm}
\begin{picture}(0.1,0.1)
\put(3.15,3.0){\epsfig{file=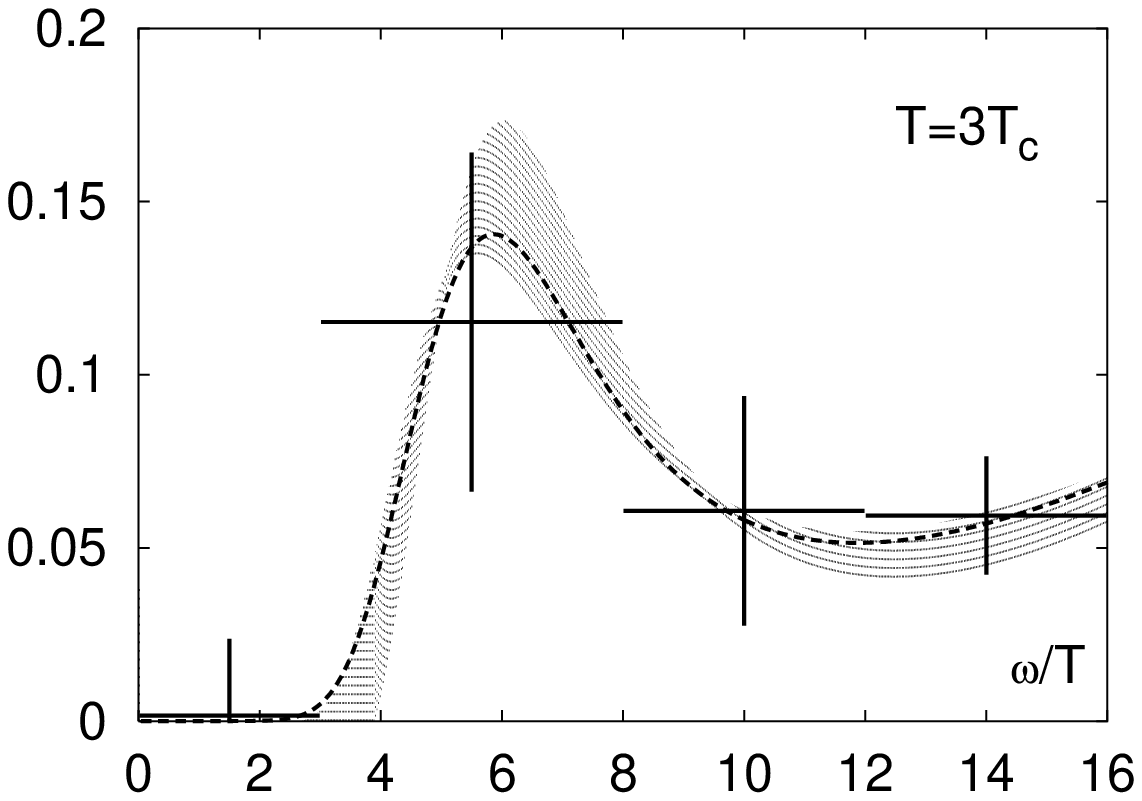,width=40mm}}
\end{picture}
\caption{
\label{fig:spectral}
Reconstructed vector spectral function $\sigma_V$ 
in units of $\omega^2$ at zero momentum (a) and the resulting zero
momentum differential dilepton rate (b) at $T/T_c = 1.5$ (doted line) and 3
(dashed line). The solid lines give the free spectral function (a) 
and the resulting Born rate (b). The insertion in (a) shows the error 
band on the spectral function at $3T_c$ obtained from a jackknife analysis 
and errors on the average value of $\sigma_V(\omega, T)/\omega^2$ in four 
energy bins (see text).}
\end{figure*}

In Fig.~\ref{fig:spectral}a we show spectral functions 
reconstructed from the vector current correlation functions at
$T/T_c = 1.5$ and 3 on the $64^3\times 16$ lattices. In the
MEM analysis we have taken into account energies up to $\omega /T =4 N_\tau$
and we used the lowest order zero temperature perturbative result 
$m(\omega)= 0.75\;\pi^{-2}\; \omega^2$ as a default model. 
In order to test the statistical significance of our results we have
split our total data sets in 8 jackknife blocks and performed a MEM
analysis on each of these blocks. The resulting jackknife error, which is 
given as a band in the insertion of Fig.~\ref{fig:spectral}a, shows that
our data sample is large enough to yield statistically significant results.
We also have checked that a variation of the default model by 20\%
only leads to minor changes in the low energy part of $\sigma_V$,
which stay within the error band shown in Fig.~\ref{fig:spectral}a.
Another question is to what extent the spectral function obtained from
the MEM analysis is unique. 
As the MEM analysis is based on a finite 
data set the reconstruction of the spectral function itself is not unique.
This uncertainty is incorporated in the MEM-error, which is calculated
from the covariance matrix of the spectral function \cite{Bryan}
for four energy bins in the interval $0 \le \omega /T \le
16$. The resulting error on the average value
of $\sigma_V(\omega, T)/\omega^2$ in these bins is also shown in 
Fig.~\ref{fig:spectral}a.

The reconstructed spectral
functions show a broad enhancement over the free spectral function 
for $\omega/T\; \gsim\; 16$, \ie $\omega a\; \gsim\; 1$. This regime clearly
is influenced by lattice cut-off effects. In fact, a similar enhancement
is observed in $T=0$ spectral functions and has been
attributed to unphysical contributions of the heavy fermion doublers
present in the Wilson (and clover) lattice fermion action 
\cite{Yamazaki}.
As discussed above this regime of large energies does not contribute
to the central values of the vector correlator for $\tau T \simeq 1/2$.
The relevant energy regime is given in the insertion of 
Fig.~\ref{fig:spectral}a. 
The enhancement observed for $G_V(1/2T,T)$ thus is due to the peak in 
$\sigma_V(\omega,T) /\omega^2$ found for $\omega/T \; \simeq\;  (5-6)$.
We furthermore note that for both temperatures the spectral functions
have a sharp cut-off at small energies. They drop
rapidly below $\omega/T \; \simeq\; 5$ and vanish below 
$\omega/T\; \simeq\;  3$. This is in contrast to perturbative 
calculations of $\sigma_V(\omega,T)$, which lead to a 
divergent spectral function for small energies.

In Fig.~\ref{fig:spectral}b we show the differential dilepton rate
calculated from Eq.~\ref{rate} and the low energy part of $\sigma_V$
shown in Fig.~\ref{fig:spectral}a. The comparison 
with the Born rate shows that for all energies $\omega/T\; \gsim\; 4$ the 
difference is less than a factor 2. 
For energies $\omega/T\; \lsim\; 3$ the dilepton rate drops rapidly and 
reflects the sharp cut-off found in the reconstructed spectral functions.
Of course, with our present 
analysis, which is based on rather small statistics, we cannot rule
out the existence of sharp resonances like the van Hove singularities
present in the HTL-resummed calculations. It also may well be 
that the broad peak found by us will sharpen with increasing 
statistics as it has been observed in related MEM studies of hadron
correlation functions at zero temperature \cite{Hatsuda}. 
Nonetheless, any further sharpening
of resonance like peaks in a certain energy interval of the spectral 
function has to be compensated
by an even closer agreement with the free spectral function in other
energy intervals due to the constraint given by Eq.~\ref{cor2T}.

\section{Conclusions}

To conclude, we find that already close to the QCD phase transition
temperature, \ie for $T=1.5 T_c$ and $3 T_c$, the vector
correlation function deviates only by less than 15\% from the leading
order perturbative result. This also is reflected in the reconstructed
spectral function which deviates by less than a factor two from the
leading order perturbative form for energies $\omega\; \gsim\; 4 T$.
The most pronounced feature of the spectral function and the resulting
dilepton rate is the presence of a sharp cut-off at low energies,
$\omega \sim (2-3)T$. If a threshold of similar magnitude persist
also closer to $T_c$, there will be no thermal contribution to the 
dilepton rate at energies $\omega\; \lsim\; 2 T_c$ during 
the entire expansion of the hot medium created in a heavy ion collision.
This is consistent with present findings at SPS energies \cite{ceres} and 
may become visible as a threshold in the dilepton rates for long-lived 
plasma states expected to be created at RHIC or LHC energies.


\noindent
{\bf Acknowledgements:} 
We thank R. Baier for very interesting discussions.
The work has been supported by the TMR network
ERBFMRX-CT-970122 and by the DFG under grant FOR 339/1-2. 
Numerical calculations have been performed on a Cray T3E
at the NIC, J\"ulich
and the APEmille at Bielefeld University.


\end{document}